\documentclass{article}
\usepackage{spconf,amsmath,amssymb,graphicx,hyperref,multirow,booktabs}

\title{ENHANCING GUIDANCE FOR MISSING DATA IN DIFFUSION-BASED SEQUENTIAL RECOMMENDATION}
\name{Qilong Yan$^{1, 2}$ \qquad Yifei Xing$^{2, 4}$ \qquad Dugang Liu$^{3}$ \qquad Jingpu Duan$^{2,*}$ \qquad Jian Yin$^{1,2,*}$\thanks{$^{*}$Corresponding authors. This work is supported by  - Mobile Information Networks National Science and Technology Major Project under Grant No.2025ZD1303200, the National Natural Science Foundation of China (U22B2060), the Key-Area Research and Development Program of Guangdong Province (2024B0101050005), and the Research Foundation of Science and Technology Plan Project of Guangzhou City (2023B01J0001, 2024B01W0004).}}

\address{$^{1}$ School of Artificial Intelligence, Sun Yat-sen University, Zhuhai, China \\
         $^{2}$ Peng Cheng Laboratory, Shenzhen, China \\
         $^{3}$ Shenzhen University, Shenzhen, China \\
         $^{4}$ University of Chinese Academy of Sciences, Beijing, China \\
         \texttt{\{yanql, duanjp\}@pcl.ac.cn, issjyin@mail.sysu.edu.cn}} 
\begin{document}
%
\maketitle
\begin{abstract}
Contemporary sequential recommendation methods are becoming more complex, shifting from classification to a diffusion-guided generative paradigm. However, the quality of guidance in the form of user information is often compromised by missing data in the observed sequences, leading to suboptimal generation quality. Existing methods address this by removing locally similar items, but overlook ``critical turning points'' in user interest, which are crucial for accurately predicting subsequent user intent. To address this, we propose a novel Counterfactual Attention Regulation Diffusion model (CARD), which focuses on amplifying the signal from key interest-turning-point items while concurrently identifying and suppressing noise within the user sequence. CARD consists of (1) a Dual-side Thompson Sampling method to identify sequences undergoing significant interest shift, and (2) a counterfactual attention mechanism for these sequences to quantify the importance of each item. In this manner, CARD provides the diffusion model with a high-quality guidance signal composed of dynamically re-weighted interaction vectors to enable effective generation. Experiments show our method works well on real-world data without being computationally expensive. Our code is available at https://github.com/yanqilong3321/CARD.
\end{abstract}
\section{Introduction}
\label{sec:intro}

Sequential recommendation aims to predict the next item of interest based on a user's historical interaction sequence \cite{kang2018self, kong2024customizing, liao2024llara}. Traditional discriminative recommendation \cite{hidasi2015session, qiu2022contrastive, wang2022unbiased} focuses on distinguishing the target item from negative samples, whereas emerging generative recommendation aims to directly generate ``oracle items'' that best match user preferences \cite{wang2023diffusion}. Compared to VAEs and GANs, which are limited in generation stability and quality, diffusion models \cite{wang2023diffusion, niu2024diffusion, yang2023generate} have emerged as the most promising technology, excelling in both stability and quality by iteratively noising and denoising items under the guidance of interaction history. However, the performance of these advanced generative models largely depends on an ideal assumption: the user's interaction history is complete. In real-world applications, missing data \cite{wang2023patch, zhang2024ssdrec} is pervasive and unpredictable, creating unreliable guidance signals that severely constrain the practical effectiveness of diffusion models.

\begin{figure}[t]
  \centering
  \includegraphics[width=\linewidth,height=6cm, keepaspectratio]{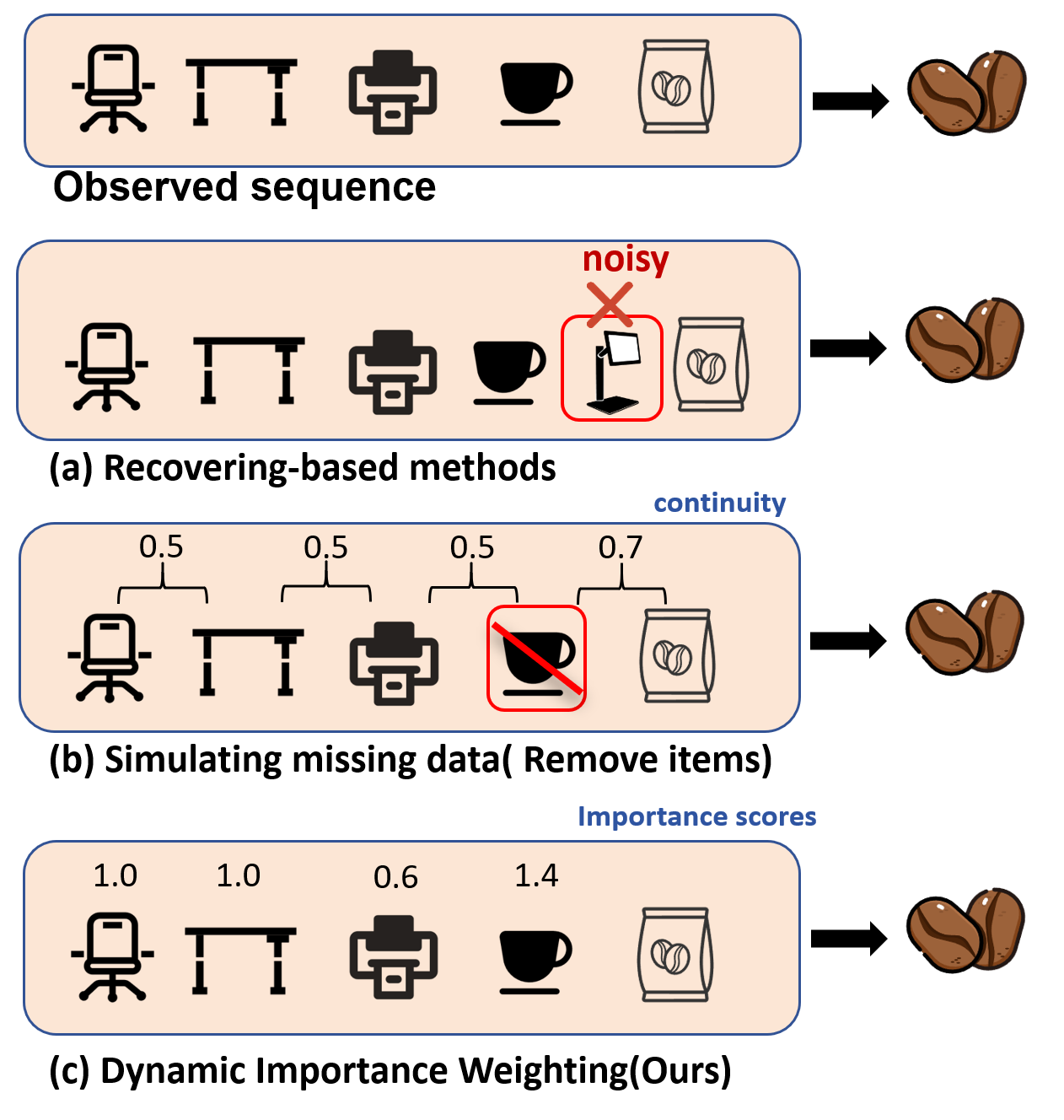}
  \caption{Comparisons of existing methods for missing data: (a) Recovering-based methods risk introducing erroneous items. (b) Simulating missingness by removing items based on local continuity. (c) Our method by re-weighting items based on their predictive importance}
  \label{fig:motivation}
\end{figure}

Two distinct approaches currently exist to address this problem: recovering-based strategies \cite{lin2023self, ma2024plug, zhang2024ssdrec} and strategies based on simulating missingness. The core idea of recovering-based strategies is to first ``repair'' or ``complete'' the incomplete sequence through interpolation \cite{hikmawati2024improve, steck2013evaluation} or augmentation before guiding the diffusion model, as seen in PDRec \cite{ma2024plug} and SSDRec \cite{zhang2024ssdrec}. Although intuitive, these methods face a fundamental challenge: the ``uncertainty'' of missing data \cite{fan2022sequential, wang2023rethinking}. Since the complete sequence as a ``ground truth'' is unavailable, the recovery process is essentially a guess, which can easily introduce new noise or distort true user preferences (as shown in Fig.~\ref{fig:motivation}(a)), potentially providing even more erroneous guidance to the diffusion model.

Strategies based on simulating missingness, in contrast, propose that rather than riskily ``recovering'' data, it is better to train the model by actively simulating missingness, enabling it to ``extrapolate'' and handle real-world missing data. For example, TDM \cite{mao2025addressing} identifies and removes highly similar items under stable preferences by calculating the ``local continuity'' between items and the ``global stability'' of the sequence, thereby constructing training samples with missingness while preserving core user preferences. However, the existing simulation mechanisms have limitations. For example, as shown in Fig.~\ref{fig:motivation}(b), due to the high similarity between ``coffee mug'' and ``coffee shop'', TDM would tend to remove one of them, believing it has minimal impact on the overall preference evolution, but this might destroy critical bridge information.

\begin{figure}[t]
  \centering
  \includegraphics[width=\linewidth]{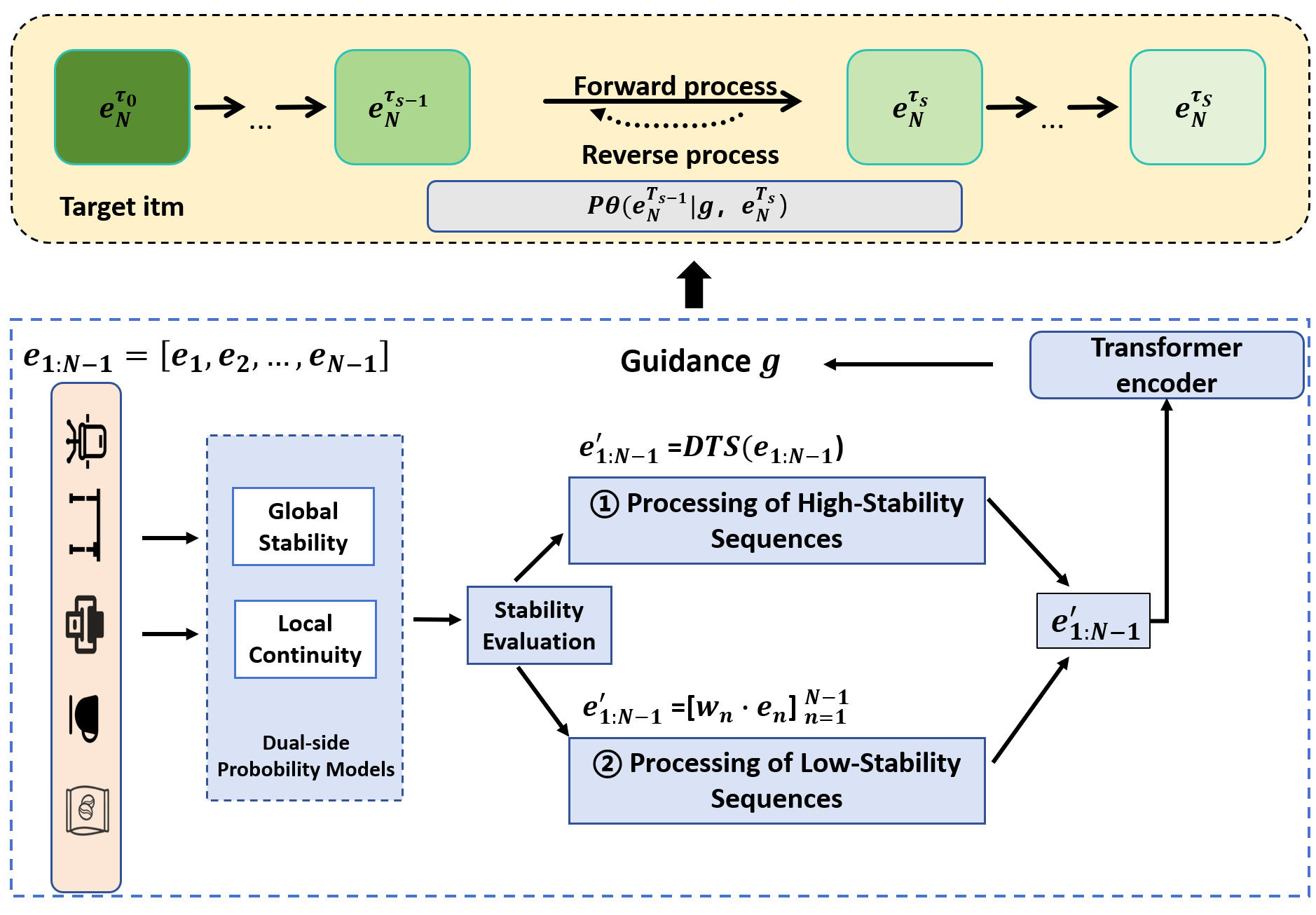}
\caption{The overall architecture follows a three-step process: (1) it first evaluates sequence stability to determine a processing path; (2) high-stability sequences are then simplified via DTS, while (3) low-stability ones are re-weighted using our counterfactual attention. The optimized sequence is then encoded to provide guidance for the diffusion model.}
  \label{fig:method}
\end{figure}
To this end, we propose CARD, a novel \textbf{C}ounterfactual \textbf{A}ttention \textbf{R}egulation \textbf{D}iffusion model. CARD employs a more robust scheme for handling sequence uncertainty. Instead of removing items, it dynamically evaluates and reshapes the importance of each item. This approach can both simulate missingness (when importance is very low) and amplify the signals of key items by increasing their attention. To achieve this, CARD introduces an attention mechanism based on counterfactual reasoning, motivated by the principle that an item's importance should be measured by its ``contribution'' to predicting future user preferences, rather than its local similarity to adjacent items. Specifically, CARD quantifies this contribution by comparing prediction error in two scenarios: one including the item's information and one excluding it (the counterfactual case). The ``Prediction Error Reduction'' brought by the item's inclusion is used to adjust its importance weight. As shown in Fig.~\ref{fig:motivation}(c), removing the ``coffee mug'' would cause a sharp increase in prediction error for subsequent items like ``coffee beans.'' Therefore, CARD will increase the attention weight of the ``coffee mug,'' identifying it as a key item that signifies a user's interest shift. Similarly, an item like a ``printer'' that might interfere with subsequent predictions would have its attention weight reduced. Finally, to enhance computational efficiency, CARD only performs attention calculations for low-stability sequences and for items that are potential turning points. For high-stability sequences, CARD adopts the mature DTS strategy, as it remains an effective means of reducing redundancy. In summary, our contributions are threefold: (1) a novel counterfactual attention mechanism that defines item importance by its predictive contribution (``Prediction Error Reduction''), rather than local similarity; (2) an efficient routing strategy that selectively applies this attention mechanism to significantly reduce computational overhead; and (3) extensive experiments validating CARD's effectiveness and efficiency on real-world datasets.

\section{METHODOLOGY}
\label{sec:method}

In this section, we first briefly introduce the preliminaries of diffusion models in recommendation, then detail CARD's components. An overview of the architecture is shown in Fig.~\ref{fig:method}.

\vspace{-1mm}

\subsection{Preliminaries: Diffusion Models for Recommendation}
A diffusion model generates a target item embedding $\mathbf{e}_N^0$ from Gaussian noise $\mathbf{e}_N^{\tau_S} \sim \mathcal{N}(0, \mathbf{I})$ through a reverse process. This process is conditioned on a guidance vector $\mathbf{g}$, which is derived from the user's historical interaction sequence $\mathbf{e}_{1:N-1}$. The training objective is to predict the original item embedding $\mathbf{e}_N^0$ from its noised version $\mathbf{e}_N^{\tau_s}$ at step $\tau_s$. During inference, we employ classifier-free guidance \cite{ho2022classifier} to combine conditional and unconditional predictions, enhancing generation quality:
\begin{equation}
    \tilde{f}_\theta(\cdot) = (1+w)f_\theta(\mathbf{e}_N^{\tau_s}, \mathbf{g}, \tau_s) - w f_\theta(\mathbf{e}_N^{\tau_s}, \mathbf{\phi}, \tau_s),
\end{equation}
where $w$ is the guidance strength and $\mathbf{\phi}$ is a dummy token for unconditional generation.

\subsection{Dynamic Guidance Optimization via Routing}
The core of CARD is its ability to create a high-quality guidance vector $\mathbf{g}$ by dynamically optimizing the user sequence based on its stability.
\subsubsection{Stability Evaluation.}
Following TDM \cite{mao2025addressing}, we first quantify the local continuity between adjacent items $\mathbf{e}_n$ and $\mathbf{e}_{n+1}$ using a softmax-normalized cosine similarity:
\begin{equation}
    \text{con}_n = \frac{\exp(\text{sim}(\mathbf{e}_n, \mathbf{e}_{n+1}))}{\sum_{n'=1}^{N-2} \exp(\text{sim}(\mathbf{e}_{n'}, \mathbf{e}_{n'+1}))}.
\end{equation}
The overall stability of a sequence is then measured by the entropy of this distribution, denoted as a stability score $s_k$, where a higher value indicates lower stability (i.e., a significant interest shift):
\begin{equation}
    s_k = -\sum_{n=1}^{N-2} \text{con}_n \log(\text{con}_n).
\end{equation}
Based on a predefined threshold $\lambda_{stb}$, we route the sequence. For high-stability sequences ($s_k \le \lambda_{stb}$), we adopt the DTS strategy from TDM \cite{mao2025addressing} to remove redundant items. Low-stability sequences ($s_k > \lambda_{stb}$) are processed as follows.
\subsubsection{Processing of Low-Stability Sequences.}
For low-stability sequences, which signal a user's interest shift, naively applying DTS can be detrimental. It may catastrophically damage the guidance signal by removing the very items that signify the interest shift. To address this critical flaw, CARD introduces its core contribution: a \textbf{counterfactual attention mechanism}.

Instead of removing items, we dynamically re-weight them. The importance of each item is not determined by local similarity, but by its causal impact on future predictions. Specifically, we quantify an item's importance by the \textbf{Prediction Error Reduction (PER)}. To achieve this, we introduce an auxiliary prediction task: for each position $n$, we predict the average embedding of the next $W$ items, denoted as $\bar{\mathbf{e}}_{>n}$. We then calculate the prediction loss under two conditions: one using the full history up to item $n$ (encoded as hidden state $\mathbf{h}_n$), and a counterfactual one using only the history up to item $n-1$ (encoded as $\mathbf{h}_{n-1}$). The PER is the difference between these two losses:
\begin{equation}
\text{PER}_n = \underbrace{\| \text{MLP}_{aux}(\mathbf{h}_{n-1}) - \bar{\mathbf{e}}_{>n} \|^2}_{\text{Loss without } \mathbf{e}_n} - \underbrace{\| \text{MLP}_{aux}(\mathbf{h}_n) - \bar{\mathbf{e}}_{>n} \|^2}_{\text{Loss with } \mathbf{e}_n}.
\end{equation}
A high PER value indicates a critical ``bridge'' item. This results in a weighted sequence:
\begin{equation}
    \tilde{\mathbf{e}}_{1:N-1} = [\mathbf{w}_n \cdot \mathbf{e}_n]_{n=1}^{N-1}, \quad \text{if } s_k > \lambda_{stb},
\end{equation}
where the weight $\mathbf{w}_n = 1 + \tanh(\text{PER}_n / T)$ is designed to amplify the signal of key turning points ($w_n > 1$) and suppress the influence of noisy or irrelevant items ($w_n < 1$). This provides a far more robust and nuanced guidance signal for the diffusion model.
\subsubsection{Guidance Vector Generation.}
Regardless of the path taken, the processed sequence $\tilde{\mathbf{e}}_{1:N-1}$ is fed into a Transformer encoder to produce the final guidance vector: $\mathbf{g} = \text{Transformer}(\tilde{\mathbf{e}}_{1:N-1})$.

\subsection{Model Training and Inference}
The denoising network $f_\theta$ is trained to predict the original item $\mathbf{e}_N^0$ from a noised version, guided by the final guidance vector $\mathbf{g}$, and optimized via a mean squared error objective:
\begin{equation}
\mathcal{L} = \mathbb{E}_{\mathbf{e}_N^0, \mathbf{g}, \epsilon, \tau_s} \left[ \left\| \mathbf{e}_N^0 - f_\theta(\mathbf{e}_N^{\tau_s}, \mathbf{g}, \tau_s) \right\|^2 \right].
\end{equation}
During inference, generation starts from random noise and iteratively denoises it using classifier-free guidance (Eq. 1). The final generated embedding is used to retrieve top-K recommendations.

\section{EXPERIMENTS}

\subsection{Experimental Setup}
\noindent
\textbf{Datasets.} We conduct experiments on two widely-used real-world datasets: Zhihu\cite{hao2021large} and KuaiRec\cite{gao2022kuairec}. Following standard preprocessing, we filter out users and items with fewer than 5 interactions and ensure all sequences have a minimum length of 3. We follow the standard data splitting and evaluation protocols as in previous works \cite{mao2025addressing, yang2023generate}.

\noindent
\textbf{Evaluation Metrics.} We adopt the standard leave-one-out strategy for evaluation, ranking the ground-truth next item against random negative samples. We report Hit Ratio (HR@20) and Normalized Discounted Cumulative Gain (NDCG@20), averaged over five runs. 

\noindent
\textbf{Baselines.} We compare CARD against three groups of models: (i) Conventional Recommenders: GRU4Rec \cite{hidasi2015session}, SASRec \cite{kang2018self}, BERT4Rec \cite{sun2019bert4rec}, and AdaRanker \cite{fan2022ada}; (ii) Generative Recommenders: DiffRec \cite{wang2023diffusion}, DreamRec \cite{yang2023generate}, and TDM \cite{mao2025addressing}; and (iii) Recovering-based Models: PDRec \cite{ma2024plug}, SSDRec \cite{zhang2024ssdrec}, STEAM \cite{lin2023self}, and DiffuASR \cite{liu2023diffusion}.

\noindent
\textbf{Implementation Details.} All models are implemented in PyTorch and trained on an NVIDIA RTX 4090 GPU. The embedding dimension is 64. We use the Adam optimizer with a learning rate tuned from $\{10^{-3}, 5\times10^{-4}, 10^{-4}\}$. For CARD, we tune the stability threshold $\lambda_{stb}$ in $[0.5, 2.0]$ and future window size $W$ in $\{1, 3, 5\}$. Baselines use their officially reported hyperparameters.

\subsection{Results}
\noindent
\textbf{Main Results.}
Table~\ref{tab:main_results} presents the overall performance comparison on the Zhihu and KuaiRec datasets. We can draw several key observations. Firstly, generative models, particularly diffusion-based approaches like TDM and DreamRec, generally surpass conventional and recovering-based methods, underscoring the strength of the generative paradigm in recommendation. Secondly, our proposed model, CARD, consistently achieves state-of-the-art performance across all metrics on both datasets. Specifically, on the challenging Zhihu dataset, CARD outperforms the strongest baselines, leading to significant relative improvements of 10.30\% in HR@20 (over TDM) and 5.06\% in NDCG@20 (over DreamRec). This substantial margin validates the effectiveness of our dynamic guidance optimization strategy. By differentiating between high- and low-stability sequences and applying tailored processing—counterfactual attention for the latter—CARD successfully captures critical interest shifts, leading to a higher quality guidance signal and superior recommendation accuracy.

\begin{table}[t]
\centering
\caption{Main performance comparison (\%). Best in \textbf{bold}, second-best \underline{underlined}. Improv. is vs. runner-up.}
\label{tab:main_results}
\resizebox{\columnwidth}{!}{%
\begin{tabular}{l|cc|cc}
\toprule
\multicolumn{1}{c|}{\multirow{2}{*}{\textbf{Methods}}} & \multicolumn{2}{c|}{\textbf{KuaiRec}} & \multicolumn{2}{c}{\textbf{Zhihu}} \\ \cmidrule(lr){2-3} \cmidrule(lr){4-5}
\multicolumn{1}{c|}{} & \textbf{HR@20} & \textbf{NDCG@20} & \textbf{HR@20} & \textbf{NDCG@20} \\ \midrule
\multicolumn{5}{l}{\textit{Conventional Recommenders}} \\
GRU4Rec & 3.32$\pm$0.11 & 1.23$\pm$0.08 & 1.78$\pm$0.12 & 0.67$\pm$0.03 \\
SASRec & 3.92$\pm$0.18 & 1.53$\pm$0.11 & 1.62$\pm$0.01 & 0.60$\pm$0.03 \\
BERT4Rec & 3.77$\pm$0.09 & 1.73$\pm$0.04 & 2.01$\pm$0.06 & 0.72$\pm$0.04 \\
AdaRanker & 4.14$\pm$0.09 & 1.89$\pm$0.05 & 1.70$\pm$0.04 & 0.61$\pm$0.02 \\ \midrule
\multicolumn{5}{l}{\textit{Recovering-based Models}} \\
STEAM & 4.98$\pm$0.05 & 2.90$\pm$0.02 & 1.75$\pm$0.02 & 0.69$\pm$0.02 \\
SSDRec & 4.19$\pm$0.08 & 3.28$\pm$0.06 & 2.03$\pm$0.06 & 0.72$\pm$0.03 \\
DiffuASR & 4.53$\pm$0.02 & 3.30$\pm$0.03 & 2.05$\pm$0.02 & 0.71$\pm$0.02 \\
PDRec & 4.42$\pm$0.03 & 3.55$\pm$0.04 & 2.10$\pm$0.03 & 0.74$\pm$0.02 \\ \midrule
\multicolumn{5}{l}{\textit{Generative Recommenders}} \\
DiffRec & 3.74$\pm$0.08 & 1.77$\pm$0.05 & 1.82$\pm$0.03 & 0.65$\pm$0.09 \\
DreamRec & 5.16$\pm$0.05 & 4.11$\pm$0.02 & 2.26$\pm$0.07 & \underline{0.79$\pm$0.01} \\
TDM & \underline{5.53$\pm$0.03} & \underline{4.82$\pm$0.04} & \underline{2.33$\pm$0.04} & 0.75$\pm$0.03 \\ \midrule
\textbf{CARD (Ours)} & \textbf{5.78$\pm$0.04} & \textbf{5.02$\pm$0.03} & \textbf{2.57$\pm$0.05} & \textbf{0.83$\pm$0.02} \\ \midrule
Improv. & 4.52\% & 4.15\% & 10.30\% & 5.06\% \\ \bottomrule
\end{tabular}%
}
\end{table}

\noindent
\textbf{Ablation Study.}
To validate the contributions of CARD's key components, we conduct an ablation study with two variants: (1) CARD w/o Routing, which applies the counterfactual attention mechanism to all sequences, regardless of their stability score; and (2) CARD w/o Attention, which disables the counterfactual attention and processes all sequences using the standard DTS strategy. As shown in Table~\ref{tab:ablation}, removing the counterfactual attention (w/o Attention) leads to a significant performance degradation, proving the effectiveness of this mechanism. Removing the routing module (w/o Routing) also impacts performance and, more critically, results in substantially higher computational costs. Their combination makes CARD both effective and efficient.

\begin{table}[h]
\centering
\caption{Ablation study results (\%) on Zhihu and KuaiRec.}
\label{tab:ablation}
\resizebox{\columnwidth}{!}{%
\begin{tabular}{l|cc|cc}
\toprule
\multicolumn{1}{c|}{\multirow{2}{*}{\textbf{Methods}}} & \multicolumn{2}{c|}{\textbf{KuaiRec}} & \multicolumn{2}{c}{\textbf{Zhihu}} \\ \cmidrule(lr){2-3} \cmidrule(lr){4-5}
\multicolumn{1}{c|}{} & \textbf{HR@20} & \textbf{NDCG@20} & \textbf{HR@20} & \textbf{NDCG@20} \\ \midrule
w/o Routing & 5.66 & 4.89 & 2.28 & 0.77 \\
w/o Attention & 5.53 & 4.82 & 2.34 & 0.75 \\ \midrule
CARD & \textbf{5.78} & \textbf{5.02} & \textbf{2.57} & \textbf{0.83} \\ \bottomrule
\end{tabular}%
}
\end{table}

\noindent
\textbf{Efficiency comparison.}
To evaluate the efficiency of our model, we compare the training time per epoch and inference time per batch against several baselines, with results shown in Table~\ref{tab:efficiency}. Notably, CARD significantly reduces training time compared to TDM—especially on KuaiRec—by selectively applying intensive counterfactual attention to low-stability sequences via our routing mechanism. During inference, CARD maintains a speed comparable to TDM as their core diffusion processes are identical. Thus, CARD enhances accuracy with minimal overhead, showcasing its practical efficiency.

\begin{table}[h]
\centering
\caption{Running time comparison of CARD with baselines.}
\label{tab:efficiency}
\resizebox{\columnwidth}{!}{%
\begin{tabular}{l|cc|cc}
\toprule
\multicolumn{1}{c|}{\multirow{2}{*}{\textbf{Methods}}} & \multicolumn{2}{c|}{\textbf{KuaiRec}} & \multicolumn{2}{c}{\textbf{Zhihu}} \\ \cmidrule(lr){2-3} \cmidrule(lr){4-5}
\multicolumn{1}{c|}{} & \textbf{Train} & \textbf{Inference} & \textbf{Train} & \textbf{Inference} \\ \midrule
SASRec & 02m 07s & 00m 08s & 00m 10s & 00m 01s \\
AdaRanker & 03m 38s & 00m 09s & 00m 14s & 00m 01s \\
DreamRec & 03m 59s & 32m 40s & 00m 14s & 01m 31s \\
TDM & 03m 30s & 00m 34s & 00m 20s & 00m 02s \\ \midrule
CARD & 02m 40s & 00m 34s & 00m 13s & 00m 02s \\ \bottomrule
\end{tabular}%
}
\end{table}
\vspace{-2mm}

\section{Conclusion}
In this paper, we addressed the critical challenge of handling missing data in diffusion-based sequential recommendation, particularly for sequences involving significant interest shifts. We proposed CARD, a novel model that dynamically optimizes the guidance signal for the diffusion process. By introducing a routing strategy to differentiate sequence stability and a counterfactual attention mechanism to adjust the weights of key items in low-stability sequences, this approach amplifies the signals of key items while suppressing potential noise. Extensive experiments demonstrated that CARD not only achieves state-of-the-art recommendation accuracy but also improves training efficiency. Our work presents a robust and efficient paradigm for enhancing the guidance quality in generative recommendation systems.

\small
\bibliographystyle{IEEEbib}
\bibliography{strings,refs}

\end{document}